\documentclass{elsart3}

\usepackage{epsfig}

\usepackage{amssymb}

\begin{document}

\begin{frontmatter}



\title{Simulation Studies of Delta-ray Backgrounds in a Compton-Scatter Transition Radiation Detector}

\author[label1,label2]{John\ F.\ Krizmanic}
\author[label3]{Michael\ L.\ Cherry}
\author[label2]{Robert\ E.\ Streitmatter}
\address[label1]{Universities Space Research Association}
\address[label2]{NASA Goddard Space Flight Center, Greenbelt, MD 20771 USA}
\address[label3]{Dept.\ of Physics \& Astronomy, Louisiana State University, Baton Rouge, LA 70803 USA}

\begin{abstract}
In order to evaluate the response to cosmic-ray nuclei of a Compton-Scatter Transition Radiation Detector in the proposed ACCESS space-based mission, a hybrid Monte Carlo simulation using GEANT3 and an external transition radiation (TR) generator routine was constructed.  This simulation was employed to study the effects of delta-ray production induced by high-energy nuclei and to maximize the ratio of TR to $\delta$-ray background. The results demonstrate the ability of a Compton-Scatter Transition Radiation Detector to measure nuclei from boron to iron up to Lorentz factors $\gamma \sim 10^5$ taking into account the steeply falling power-law cosmic ray spectra.
\end{abstract}

\end{frontmatter}

\section{Introduction}
\label{Intro}
The proposed Advanced Cosmic-ray Composition Experiment for Space Science (ACCESS) is a dedicated space-based mission to perform direct nuclear composition measurements up to near the `knee' of the cosmic-ray spectrum, $E \sim 10^{6}$ GeV/nucleus \cite{ACCESS}.  One configuration \cite{Cherry1} employs a nearly 6 meter$^3$ Compton-Scatter Transition Radiation Detector (CSTRD) in a cubic geometry with a smaller calorimeter underneath.  The absolute size of the ACCESS instrument is driven by the requirement to obtain significant event statistics in this flux-poor energy regime while being constrained by the achievable spacecraft size and mass.

The Compton-scatter technique exploits the facts that 1) the x-ray spectra of emitted  TR can be hardened by a judicious choice of radiator material, thickness, and spacing and 2) the greatest energy dependence of the TR is at high frequencies \cite{Cherry1}.
For a radiator with thickness $l_1$, plasma frequency $\omega_1$, and spacing $l_2$, interference effects from the superposition of amplitudes from each interface yield pronounced maxima and minimum in the TR spectra.  The highest frequency maximum is at $\omega_{max} = \frac{l_1 \omega_1^2}{2 \pi c}(1+\rho)$ while the saturation energy is $\gamma_s \approx \frac{0.6 \omega_1}{c}\sqrt{l_1 l_2 (1 + \rho)}$, where $\rho =0$ except for materials with complex dielectric constants (i.e., metals) where $\rho = 1$.  (In order to be conservative, the simulations employed in this study assume $\rho = 0$).  Thus an ensemble of aluminum foil radiators with $100 - 200$ $\mu$m thickness separated by several mm leads to $\gamma_s \approx 10^5$ and $\hbar \omega_{max} \approx $ 130 keV (for $\rho=0$). As the Compton-scattering cross-section is significant at these energies, these photons will scatter away from the initiating particle's trajectory and thus spatially separate the TR signal from the ionization path.  Furthermore, the significant flux of x-rays with energies $> 50$ keV can be detected with an inorganic (e.g., CsI) scintillator.  
The principle of Compton-scatter TRDs has been demonstrated in accelerator test beams \cite{Cherry2}.

The thick radiators used to produce such hard x-rays also induce $\delta$-ray production.  The impact of this background is exacerbated in cosmic-ray measurements by the vast number of sub-TR events resulting from the power-law nature of the cosmic-ray spectrum.  The study presented in this paper optimizes the CSTRD signal in relation to this background.

\section{Simulation Studies}
\label{Simu}
A hybrid Monte Carlo simulation has been constructed using GEANT3 \cite{GEANT3} interfaced to an external TR generator routine based on the formalism of Ter-Mikaelian \cite{Cherry3}.  GEANT3 models the dominant electromagnetic processes in the energy range 10 keV to 10 TeV with the cross-sections smoothly extending to higher energies. For heavy ions, ionization losses are simulated along with the discrete process of $\delta$-ray production.  Hadronic interactions are not simulated for heavy ions in GEANT3,
and the ACCESS calorimeter and ancillary charge measurement detectors are not modeled in this study. 

Although multiple CSTRD configurations were investigated, the results employing a radiator arrangement with the highest TR saturation Lorentz factor are presented.  The configuration consists of a large radiator stack surrounded on all six sides by highly segmented scintillators which measure both the Compton-scattered TR and ionization losses, thus providing the primary particle charge determination.  The radiator consists of 150 Al foils with 150 $\mu$m thickness separated by 1 cm vacuum gaps with an areal size of $160 \times 160$ cm$^2$.  The radiator stack is surrounded by six $160 \times 160$ cm$^2$ planes of CsI scintillator in a cubic geometry.  Each scintillator module contains a layer of 2 mm thick CsI, 250 $\mu$m silicon PIN diode readout, and 5 mm of G10 encapsulated in a 1 mm Al housing.  The CsI is segmented into $2.5 \times 2.5$ cm$^2$ pixels with each pixel recording the deposited energy on an event-by-event basis with a 50 keV threshold energy.

The pixel TR energy distribution peaks at an energy per pixel near 100 keV with nearly 100\% of the Compton-scattered TR signal being recorded below 500 keV. The $\delta$-ray pixel energy distribution shows a low energy peak below 100 keV with a pronounced tail extending well above 1 MeV.  A pixel energy selection of $50 \le E_{pixel}  < 500$ keV retains virtually 100\% of the TR hit pixels while rejecting $> 40\%$ of the $\delta$-ray hits. The pixels recording the ionization of the primaries and their nearest neighbors are excluded in this pixel counting.

\begin{table}
{Table 1. \hspace{4pt} Response of a CSTRD to normally incident carbon nuclei with Lorentz factors above TR saturation. The first number in each column represents the number of pixel hits per event with $50~{\rm keV} \le E_{pixel} < 500~{\rm keV}$ while the second number (in parentheses) gives the pixel hits per event without energy selection.\medskip} 
\centering
%
%
\begin{tabular}{ccccc}
CsI layer & TR only & $\delta$-rays only & Sum & TR $+ \delta$-rays  \\
\noalign{\smallskip}\hline\noalign{\smallskip}
Top & 6.7 (7.9) & 0.5 (0.7) & 7.1 (8.6) & 7.1 (8.6) \\
Sides & 23.8 (27.9) & 1.5 (2.3) & 25.3 (30.2) & 26.3 (31.4) \\
Bottom & 18.3 (22.3) & 1.7 (4.5) & 20.0 (26.8) & 19.7 (26.3) \\
\hline
Total & 48.8 (58.1) & 3.7 (7.6) & 52.4 (65.7) & 53.1 (66.3) \\ 
\end{tabular}
\end{table}

\begin{figure*}[t]

\begin{center}
\psfig{file=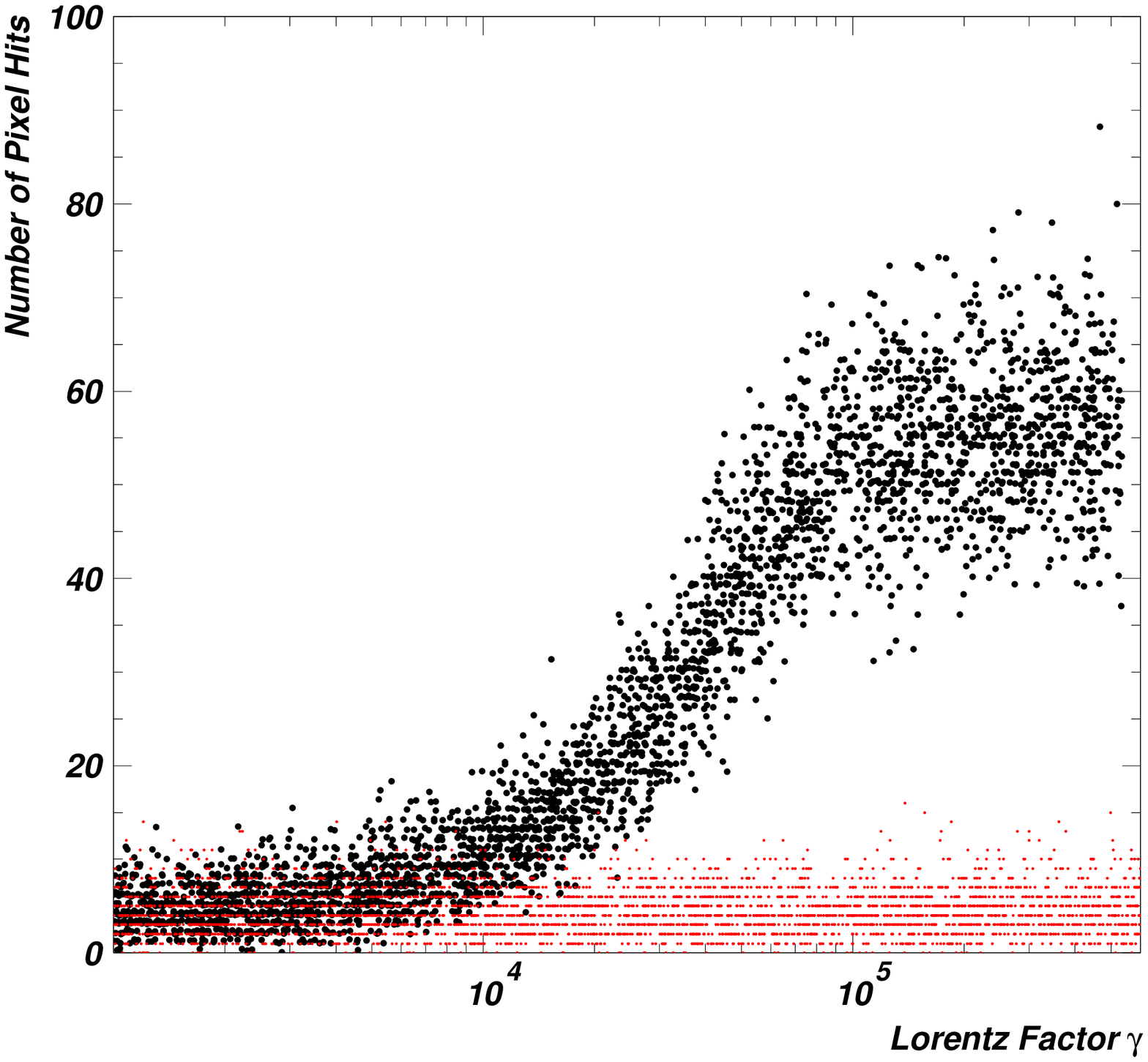,height=7.5cm}
\hspace{0.5cm}
\psfig{file=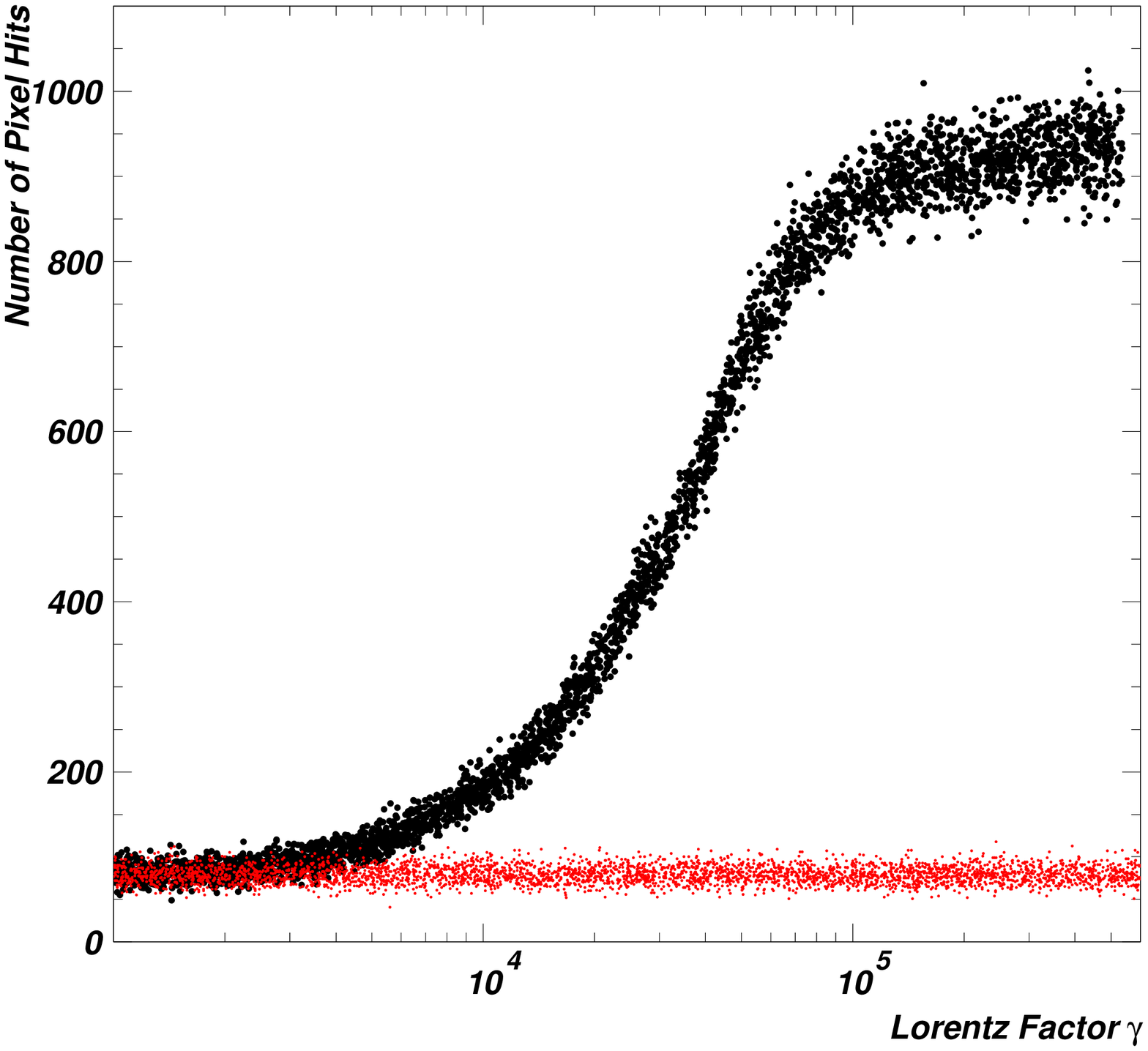,height=7.5cm}
\end{center}
\caption{The simulated total number of pixels hit for normally incident carbon (left) and iron (right) nuclei.  The response to the background $\delta$-ray process alone is illustrated by the lower, horizontal distribution.}
\end{figure*}

Table 1 details the response for normally incident carbon nuclei with $\gamma > 10^{5}$ for simulated TR only, $\delta$-ray background only, and signal + background.  Note that the pixel occupancy is at such a low level that summing the individual signal and background samples equals the result for simultaneously simulating the two processes; i.e., there are very few events in which a $\delta$-ray and TR photon hit the same pixel.  The results indicate that the signal is $48.79/3.65 = 13.4$ times larger than the background.  By varying the scintillator thickness from 1 to 8 mm, the signal/background ratio was maximized at 2 mm CsI thickness.   Furthermore, a substantial reduction in the $\delta$-ray background was obtained by placing the scintillators on the outside of the radiator stack versus interspersing them within the stack.  No significant gain in the signal/background ratio was obtained by rejecting pixels further from the trajectory of the primary than just the hit pixel plus nearest neighbors.  

These simulations were performed using nuclear primaries from boron to iron and incident energies chosen from a flat $\log{E}$ distribution to enhance the measured high energy response. Figure 1 shows the results for carbon and iron for TR + background and for $\delta$-ray background alone.  This radiator configuration yielded a mean saturated value of $\sim960$ TR photons for carbon and $\sim18,000$ TR photons for incident iron nuclei with, on average, $\sim 27\%~(20\%)$ of the TR photons at energies greater than 50 (100) keV .

The effective dynamic range of the TR measurement is obtained from a statistical analysis. Assuming a power law spectrum $\varphi(E) \sim E^{-\alpha}$ and a TR threshold energy of $E_{thr}$, the number of events above the TR threshold goes as $N_{signal} \sim E_{thr}^{1-\alpha}$.  Some number of sub-threshold events accompanied by $\delta$-rays will masquerade as high-energy events with TR. If the number of these misidentified low-energy events is required to be no more than $\beta N_{signal}$, then the probability of a low-energy event with $\delta$-rays fluctuating upward to mimic the signal of a valid event above $ E_{thr}$ must be no more than 
$P = \beta(E_0/E_{thr})^{\alpha-1}$.  Using the conservative values of $\alpha=2.75$, 
$\beta = 0.01$, $E_0 = 1$ GeV, and $E_{thr} = 10^5$ GeV, the probability is evaluated as approximately $2 \times 10^{-11}$ or $\sim 7~\sigma$ assuming a single-sided Gaussian probability distribution function.  Thus the low end of the TR dynamic range is defined as that energy where the TR response is $7~\sigma$ above the average value of the sub-TR response.
On a power-law decreasing spectrum, the flux of particles at the high end of the energy range is depleted as compared to lower energies. Thus,
the high end of the TR dynamic range can have a more modest $2~\sigma$ separation between the signal and the saturated TR value.

\begin{table}
{Table 2. \hspace{4pt} Dynamic range of TR measurements for various incident nuclei. \medskip} 
\centering
%
%
\begin{tabular}{ccccc}
\hline\noalign{\smallskip}
Nuclei & Effective $\gamma_{low}$ & Effective $\gamma_{high}$ & Dynamic Range \\
\noalign{\smallskip}\hline\noalign{\smallskip}
$^{11}$Boron & $2.4 \times 10^4$ & $4.5 \times 10^4$ & 1.9 \\
$^{12}$Carbon & $2.2 \times 10^4$ & $5.2 \times 10^4$ & 2.4 \\
$^{16}$Oxygen & $1.6 \times 10^4$ & $5.8 \times 10^4$ & 3.6 \\
$^{28}$Silicon & $1.1 \times 10^4$ & $7.2 \times 10^4$ & 6.5 \\
$^{56}$Iron & $7 \times 10^3$ & $8.3 \times 10^4$ & 11.9 \\
\end{tabular}
\end{table}

Table 2 details the effective dynamic ranges for various nuclear species based upon the results of this analysis. As the data demonstrate, the modeled CSTRD can measure the Lorentz factors of incident nuclei from boron to iron.  Furthemore, the range of Lorentz factors that guarantee signal separation from the sub-TR background and the saturated value significantly increases as the atomic number of the incident nuclei increases.  

Although hadronic interactions of the incident nuclei are not modeled for this study, an estimate of the attenuating effects of the relatively thick radiators and CsI scintillators in the CSTRD can be obtained by consideration of the nuclear inelastic cross-sections.  For iron primaries incident upon an aluminum target, the nuclear interaction length is given as 20.3 g/cm$^2$ \cite{Sihver} while the grammage of the CSTRD modeled in this study is 11.5 g/cm$^2$.  Thus, approximately 43\% of an incident iron flux will be lost due to hadronic inelastic collisions. Events with charge-changing interactions will be recognized by observing a difference in the charge measurement of the particles entering versus exiting the detector.

\section{Conclusions}
A Monte Carlo simulation has been employed to simulate the response of a Compton-scatter transition radiation detector and to optimize the design with respect to the background process of $\delta$-ray production.  The results indicate that a space-based CSTRD can perform nuclear composition measurements up to a Lorentz factor of $10^5$.  However, while the dynamic range is more than a factor of 10 for cosmic-ray iron measurements, the range is more limited for lower-$Z$ nuclei due to the relative reduction in TR generation.




\vspace{1.cm}
\begin{thebibliography}{00}
\bibitem{ACCESS} T.L.\ Wilson \& J.P.\ Wefel, NASA report TP-1999-209202 (1999);  
M.H.\ Israel et al., NASA NP-2000-05-056-GSFC (2000)
\bibitem{Cherry1} M.L.\ Cherry \& G.L.\ Case, Astropart.Phys. \textbf{18}, 629 (2003)
\bibitem{Cherry2} G.L.\ Case et al., NIM \textbf{A524}, 257 (2004)
\bibitem{GEANT3}  CERN Program Library Long Writeup W5013 (1993)
\bibitem{Cherry3} M.L.\ Cherry, Phys. Rev. \textbf{D10}, 2245 (1978)
\bibitem{Sihver} L. Sihver et al., Phys Rev C47, 1225 (1993)
\end{thebibliography}
\end{document}